\documentclass[epj]{svjour}
\usepackage[utf8]{inputenc}
\usepackage{latexsym}
\usepackage{graphics}
\usepackage{xcolor}
\usepackage{color}
\usepackage{amsfonts}
\usepackage[T1]{fontenc}
\usepackage{graphics}
\usepackage[]{graphicx}
\usepackage{latexsym}
\usepackage[figuresright]{rotating}
\usepackage{float}
\usepackage{amssymb}
\usepackage{amsmath}
\usepackage{lgreek}
\begin{document}
\title{Optical tweezers in a dusty universe}
\subtitle{Modeling optical forces for space tweezers applications}
\author{P. Polimeno\inst{1,2} \and A. Magazz\`{u}\inst{1} \and M. A. Iat\`{\i}\inst{1} \and R. Saija\inst{2}\thanks{\emph{Email:} rsaija@unime.it} \and L. Folco\inst{3,4} \and D. Bronte Ciriza\inst{1} \and M. G. Donato\inst{1} \and  A. Foti\inst{1} \and P. G. Gucciardi\inst{1} \and A. Saidi\inst{2} \and C. Cecchi-Pestellini\inst{5} \and A. Jimenez Escobar\inst{5} \and E. Ammannito\inst{6} \and G. Sindoni\inst{6} \and I. Bertini\inst{7,8} \and V. Della Corte\inst{8} \and L. Inno\inst{7} \and A. Ciaravella\inst{5}\thanks{\emph{Email:} angela.ciaravella@inaf.it} \and A. Rotundi\inst{7,8}\thanks{\emph{Email:} rotundi@uniparthenope.it} \and O. M. Marag\`{o}\inst{1}\thanks{\emph{Email:} onofrio.marago@cnr.it}
%
}                     
%
%
\institute{CNR-IPCF, Istituto per i Processi Chimico-Fisici, I-98158, Messina, Italy \and
Dipartimento di Scienze Matematiche e Informatiche, Scienze Fisiche e Scienze della Terra, I-98166,
Universit\`{a} di Messina, Italy \and Dipartimento di Scienze della Terra, Universit\`{a} di Pisa,
Pisa, Italy \and CISUP, Centro per l'Integrazione della Strumentazione dell'Universit\`{a} di Pisa,
Pisa, Italy \and INAF Osservatorio Astronomico di Palermo, Palermo, Italy \and ASI Agenzia Spaziale
Italiana, Roma, Italy \and Dipartimento di Scienze e Tecnologie, Universit\`{a} di Napoli
"Parthenope", Napoli, Italy \and INAF-IAPS, Istituto di Astrofisica e Planetologia Spaziali, Roma,
Italy }
\date{Received: date / Revised version: date}
%
\abstract{Optical tweezers are powerful tools based on focused laser beams. They are able to trap,
manipulate and investigate a wide range of microscopic and nanoscopic particles in different media,
such as liquids, air, and vacuum. Key applications of this contactless technique have been
developed in many fields. Despite this progress, optical trapping applications to planetary
exploration is still to be developed. Here we describe how optical tweezers can be used to trap and
characterize extraterrestrial particulate matter. In particular, we exploit light scattering theory
in the T-matrix formalism to calculate radiation pressure and optical trapping properties of a
variety of complex particles of astrophysical interest. Our results open perspectives in the
investigation of extraterrestrial particles on our planet, in controlled laboratory experiments,
aiming for \textit{space tweezers} applications: optical tweezers used to trap and characterize
dust particles in space or on planetary bodies surface.
} 
\maketitle

\section{Optical tweezers}
Mechanical effects of light were first argued by Kepler to explain comet tails
\cite{kepler1619cometis}. The advent of the laser technology in the 60s enabled to reproduce these
effects on our planet starting a real scientific revolution
\cite{ashkin1970acceleration,ashkin1970atomic}. Optical tweezers
\cite{ashkin1986observation,jones2015optical,marago2013optical,Polimeno2018} are tools based on
tightly focused laser beams capable to trap, manipulate, and characterize a wide range of
microscopic and nanoscopic particles, in liquids, air, and vacuum
\cite{marago2013optical,Polimeno2018}. Since the pioneering work by Ashkin
\cite{ashkin1970acceleration,ashkin1970atomic,ashkin1986observation}, that led him to the Nobel
prize in Physics 2018, key applications of this contactless manipulation technique have been
developed in a wide range of fields: from biology, soft matter, and ultra-sensitive spectroscopy to
atomic physics, nanoscience, photonics, spectroscopy, and aerosols science \cite{jones2015optical}.
A crucial advancement has been the realization of Raman tweezers, \textit{i.e.}, the coupling of
optical tweezers with a Raman spectrometer \cite{petrov2007raman}. This allows the chemical and
physical analysis of a trapped particle through its vibrational fingerprints
\cite{gillibert2019raman,donato2018optical}.

Despite this tremendous experimental progress, the accurate light scattering modeling of optical
tweezers, that takes into account particle size, shape, and composition, has been developed only
recently \cite{Polimeno2018,borghese2007optical,borghese2008radiation}. In the limiting cases of
particles much smaller or much larger than the laser wavelength, optical forces in optical tweezers
can be divided in two components \cite{jones2015optical}: a gradient force, proportional to the
intensity gradient of the laser spot, responsible for trapping; and a scattering force,
proportional to the light intensity that tends to push particles away from the laser focus
destabilizing single-beam trapping of particles with large extinction. Such detrimental effects can
be suppressed through the use of two counter-propagating beams to null the opposite scattering
forces \cite{zemanek2003theoretical}. These dual-beam traps are based on the use of low numerical
aperture (NA) lenses and allow the trapping of particles with reduced incident power in a focal
region that is wider than for standard optical tweezers \cite{donato2018optical}. Thus they are
well suited for operation in air or vacuum where detrimental effects by radiation pressure are
enhanced by the reduced viscous damping \cite{gong2018optical,svak2018transverse,alali2020laser}.

\begin{figure}
  \centering
  \resizebox{12 cm}{!}{%
  \includegraphics{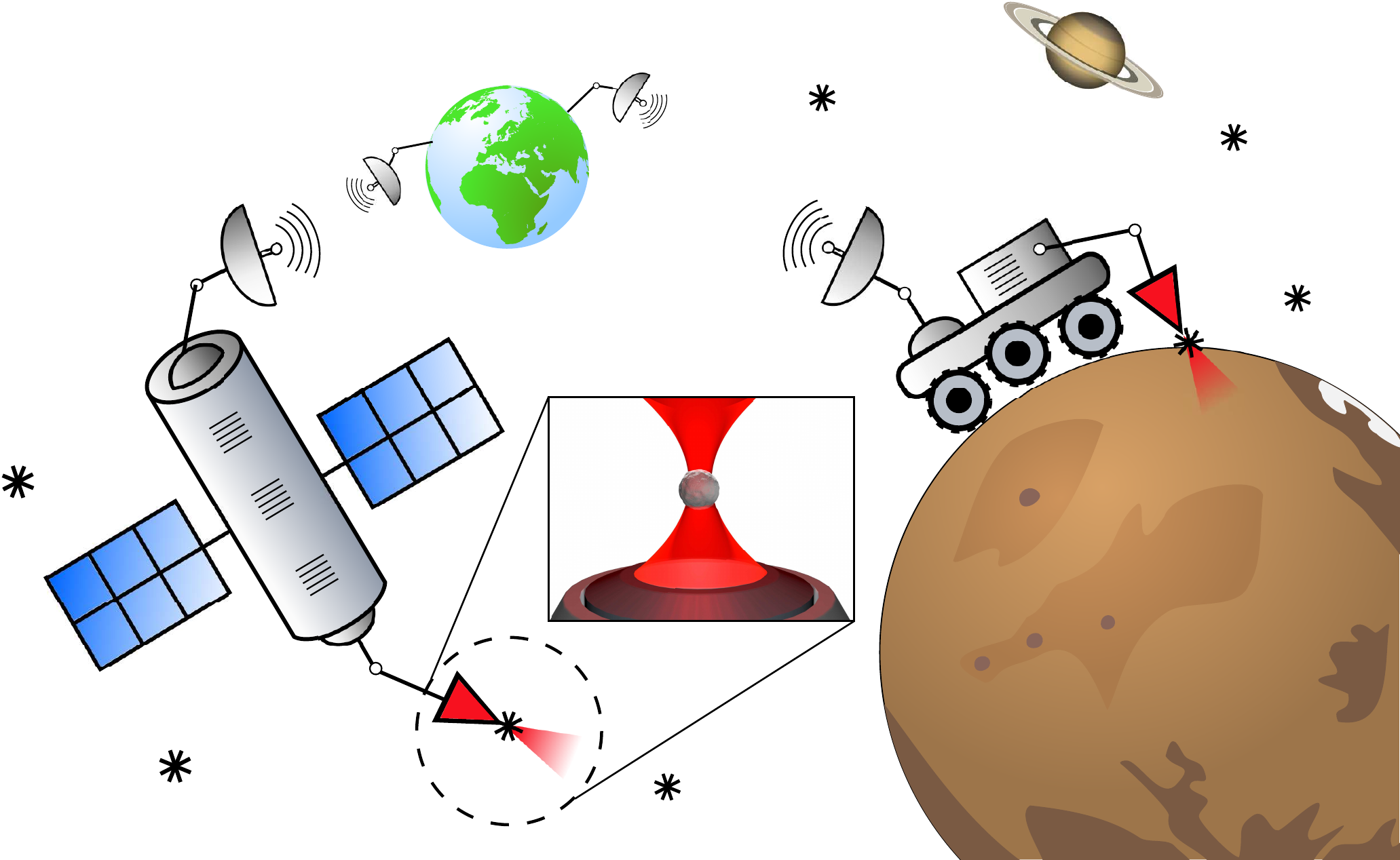}
  }
  \caption{Pictorial representation of space tweezers, space applications of optical tweezers. Interplanetary or planetary dust can be collected and investigated  directly \textit{in situ} (open space or extraterrestrial surfaces). The inset represents a closeup up of a grain of interplanetary dust trapped by a single-beam optical tweezers.}\label{figure1a}
\end{figure}

Notwithstanding the remarkable improvement in optical trapping techniques, their application to
planetary exploration (see the pictorial representation in Fig. \ref{figure1a}) is still to be
developed, even though already conceived by, \textit{e.g.}, NASA \cite{NASA2014}. The development
of the optical trapping technique to collect and analyze \textit{in situ} or return to Earth a
variety of extraterrestrial particles will open doors to information on space materials that is
currently unreachable: 1) as the dust volatile component, not measurable in situ by dust
instruments, \textit{e.g.}, those onboard Rosetta/ESA, and not retrievable by samples return
missions, \textit{e.g.}, Stardust/NASA, Hayabusa; 2) have biases due to collection media
contamination, \textit{e.g.}, aerogel used by the Stardust/NASA space probe
\cite{rotundi2008combined,rotundi2014two}).

Here, we first review the role of dust in the universe: from interstellar, to interplanetary,
cometary and planetary dust particles. Thus, we describe the models and methods we use to calculate
light pressure and optical trapping properties on a variety of realistic dust particle models.
Then, we show results on solar radiation pressure calculations that can help to a greater
understanding of micro-particle dynamics as well as to estimate its detrimental role in optical
trapping in space. Finally, we consider and compare results of calculations for optical trapping of
dust particles in standard optical tweezers in water (typical laboratory conditions) with those
calculated for OT in air or in space.

\section{A dusty universe}
We live in a dusty galaxy, one of the billions of galaxies in the universe. Almost all of them are
dusty. Up to the mid-20th century, all of this dust was considered as an unfortunate impediment to
making precise observations of stars and galaxies, regarded as the most important items in the
universe. The modern view of dust as a cosmic component is almost entirely the reverse of that
earlier view. We now know that almost every aspect of the formation of planets, stars, and galaxies
is influenced in some way by interstellar dust \cite{CCP12}. The involvement of dust grains in
providing molecules important for the origin of life, and in the safe transmission of those species
to newly-forming planets orbiting Sun-like stars \cite{McGuire}, has been one of the greatest
surprises of all. It is no wonder that considerable efforts are currently placed in trying to
understand nature, composition, and evolution of dust in the interstellar medium of galaxies.

There is now a robust evidence that dust grains condense in circumstellar environments, mainly AGB
(Asymptotic Giant Branch) stars \cite{gail2014} and supernovae \cite{sarangi15,gall14}, from which
they are ejected in the general interstellar medium. There, dust is found to be tightly mixed with
the gas, with the dust representing only a minor fraction of the total mass. The gas-to-dust mass
ratio locally assumes the fairly constant value of 100 \cite{giannetti17}. The dependence of this
ratio on the metal content among galaxies and within a galaxy is an important issue from a cosmic
perspective for a number of reasons \cite{zafar13}, including the appearance of the first solids in
the early universe \cite{gall11}. In late-type galaxies, like our own,  this ratio scales with the
metallicity,  decreasing with the galactocentric radius, a clear indication that dust growth in the
interstellar medium dominates over destruction \cite{mattsson14}. As a consequence, stardust,
literally dust formed around stars, is not the same as interstellar dust, and must be regarded as
the raw material from which true interstellar grains are formed. Such a material is modified and
destructured by violent interstellar processes \cite{Bocchio14}, before being reformed and
reassembled in denser interstellar regions \cite{Draine09}.

Modifying processes continue to act throughout the lifetime of a dust grain, some hundreds of
millions of years in the Milky Way galaxy. In some cases, such modifications may be catastrophic,
as during the formation of a planetary system, a filter that severely modifies the composition of
interstellar dust. Those materials passing through the filter are the most robust to destruction
and erosion. In the solar system, the patterns of isotopic abundances that are found in survived
presolar grains clearly identify their origins in the cool envelopes of evolved low mass stars and
in supernovae ejecta. These particles have passed through many episodes of possible destruction,
including their ejection from the stellar envelopes, their passage through the interstellar medium,
where they have been subjected to intense radiation fields and dynamical shocks, their
incorporation into the molecular cloud that formed the solar system, and through all the varied and
violent processes involved in the formation of the Sun and its planets. As near-stellar dust is
modified in the interstellar medium and becomes interstellar dust, so interstellar dust may be
severely  modified, when not obliterated, when it is incorporated in the gas that forms newly-born
planetary system.

\begin{figure}
  \centering
  \resizebox{\textwidth}{!}{%
  \includegraphics{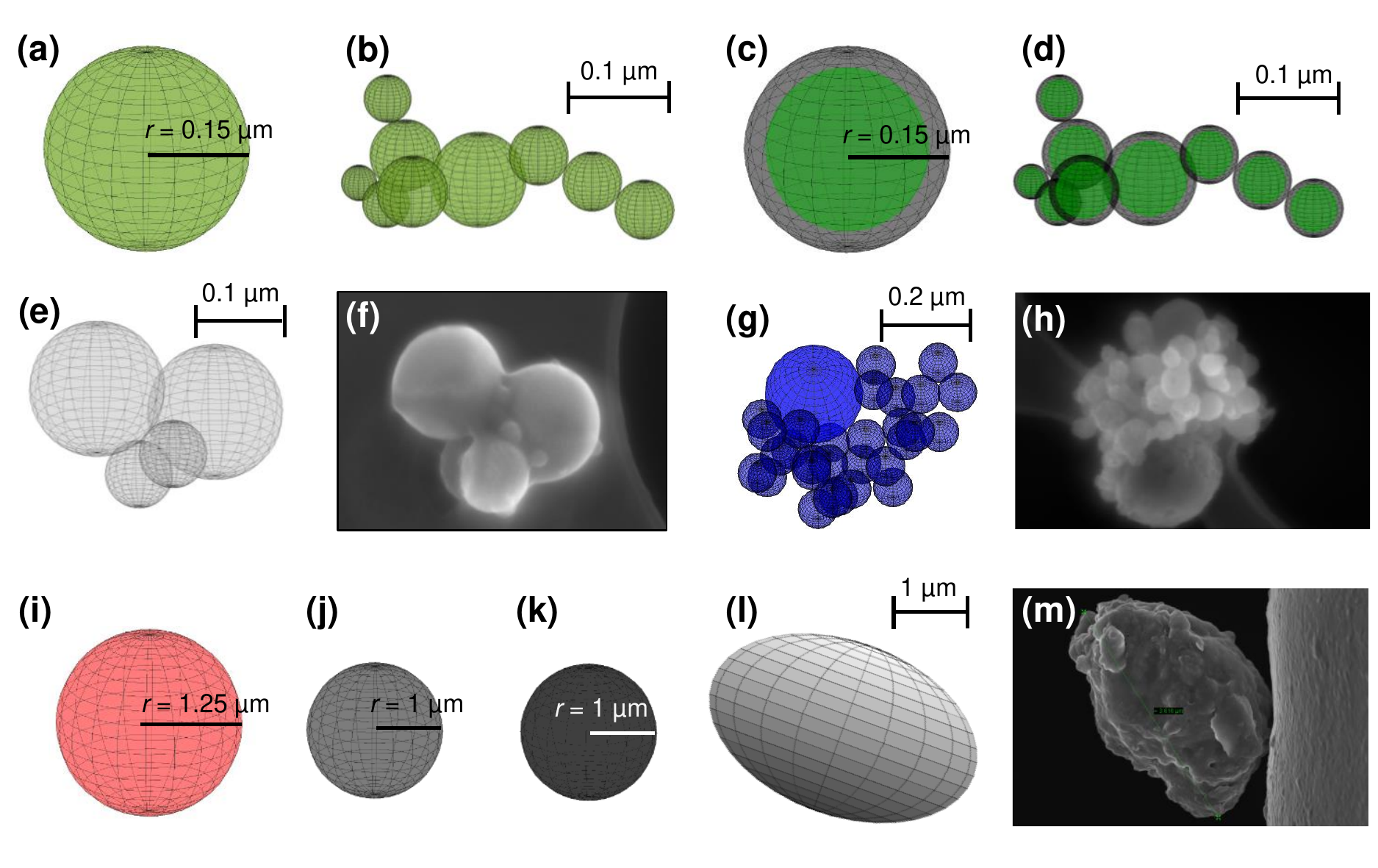}
  }
  \caption{Scattering models with shape and composition inspired by interstellar, interplanetary
  (DUSTER mission \cite{rietmeijer2016laboratory}), and planetary \cite{oliva2019database} particles.
  On the top, the models emulate hypothetical interstellar dust grains whose constituents are olivine and aliphatic carbon.
  In (a) and (b), the constituents refractive indexes are mixed according to the Bruggeman criterion. Instead, in (c) and (d),
  the olivine is considered covered by a carbon layer. In (e), the model of a silica particle arranged in quenched melt spheres shown in the
  FESEM image \cite{rietmeijer2016laboratory} (f). In (g), the model of condensed Ca[O] nanograins that are accreted onto a larger melted aggregate
  of tiny carbonate grains shown in the FESEM image \cite{rietmeijer2016laboratory} (h). The larger sphere is calcite and the other spheres
  are CaO. In (i), a spherical model of the particle Fe, Mg-rich 'TP2' \cite{rotundi2014two}, in which we consider an effective refractive
  index constructed mixing iron (67\%) and magnesium (33\%), according to the Bruggeman criterion. In (j), a spherical model of
  Martian hematite \cite{oliva2019database}. In (k), a spherical model of Lunar regolith \cite{egan1973optical}. In (l), a model of an
  ellipsoidal fassaite shown in the FESEM image (m). Here, we consider an effective refractive index constructed mixing
  silica (53\%), CaO (27\%), FeO (10\%), Al$_2$O$_3$ (10\%), according to the Bruggeman criterion \cite{bohren2008absorption}.}\label{figure1}
\end{figure}

\subsection{Interstellar dust}
Most of the information that we have about interstellar dust is obtained remotely, by the influence
of dust on various kinds of astronomical observations. These observations may be carried out at a
very wide range of wavelengths, from X-rays to radio waves, but traditionally the most important
and defining data have come from the infrared, visible, and ultraviolet parts of the spectrum,
summarized through a wavelength-dependent extinction curve along the lines of sight to individual
stars.

There are detailed extinction measurements along hundreds of lines of sight in the Milky Way Galaxy
\cite{Fitzpatrick19}, and less accurate extinction data for the interstellar media of external
galaxies \cite{Naveen20}. Such observed extinction profiles are different, but are clearly members
of the same family showing a basic similarity in their shapes: the extinction typically increases
from low values in the infrared to high values in the far ultraviolet, a near-linear portion in the
optical region, (in most cases) a pronounced and broad "bump" near 217.5 nm, and a final rise of
varying slope into the far ultraviolet. Both the general aspect of the extinction and the details
of specific curves along particular lines of sight provide useful information, and generally
indicate that grains of a wide range of sizes (roughly nanometers to micrometers) are required
\cite{CCP15}.

The dominant feature  in the extinction curve is the prominent bump in the near ultraviolet. Its
central position is fixed, although the width of the feature can vary significantly from one line
of sight to another. It is widely attributed to $\pi^\star \leftarrow \pi$ transitions in aromatic
carbon solids \cite{gadallah11} or polycyclic aromatic hydrocarbons \cite{tielensbook}. The
far-ultraviolet rise may be decomposed in a linear contribution, due to nano-sized particles, and a
non-linear component belonging to the partially invisible (because located beyond the Lyman
continuum) $\sigma^\star \leftarrow \sigma$ resonance in aromatic carbon. In the near-infrared,
there is a weak absorption feature at a wavelength of 3.4~$\mu$m, detectable on long paths through
diffuse gas. It is characteristic of absorption in the $sp^3$ (aliphatic) C-H stretching bond
\cite{Sandford91}. Further into the infrared are two stronger absorption features, at 9.7 and
18~$\mu$m, ascribable to silicate materials, from Si-O stretching and O-Si-O bending modes,
respectively. Taken together, these features strongly support the presence of some form of
carbon/hydrocarbon and silicate in the dust. X-ray scattering and absorption edges provide
constraints on grain size and composition, specifically O, Mg, Si, Fe, and C atoms \cite{Draine03}.
There is also a set of detected infrared emission features occurring at 3.3, 6.2, 7.7, 8.6, and
11.3~$\mu$m, indicative of aromatic CH groups. The mechanism responsible for the excitation of such
infrared emission may involve non-equilibrium emission from polycyclic aromatic hydrocarbons
stochastically heated to high temperatures by the absorption of individual photons from the
interstellar radiation field \cite{Leger84}. However, the requirement on microscopic sizes can be
relaxed if the emitters of the 3.3~$\mu$m and other infrared bands are heated by the chemical
energy released from reactions within larger carbon interstellar grains of mixed $sp^2/sp^3$ carbon
composition \cite{DW11}. Such structures have been in fact observed in some extragalactic objects
\cite{dartois07}. Exploiting a ternary phase diagram where the hydrogen content and the two main
bonding types ($sp^2$ and $sp^3$) for carbon constitute the poles, Dartois et al. (2007)
\cite{dartois07} were able to identify the carrier of the spectral features as an interstellar
hydrocarbon belonging to the class of polymeric-like hydrogenated amorphous carbon (a-C:H),
dominated by an aliphatic/olefinic backbone structure. The change from aliphatic to aromatic
structures may occur in environments that selectively dehydrogenate the a-C:H, providing an
opportunity for aromatic molecules to form. These observations, together with observations of very
evolved stars (protoplanetary and planetary nebulae), suggest an evolution in which aliphatics are
converted into aromatic structures \cite{antonia2008stratified,CCP10}.

The interpretation of dust observations must take also account of the available abundances along
the line of sight. The hydrogen abundance is often well-determined; abundances of other elements
relative to hydrogen are assumed, using solar or other relative values as a standard. It is unclear
which of these standards is the appropriate one to use \cite{Nieva08}, but whichever one is
adopted, the inventory of some materials is demonstrably incomplete. In fact, silicates alone are
unable to account for the entirety of either the oxygen solid phase or Fe abundances
\cite{Jenkins09}. The remainder of the Fe could be in iron oxides or in metallic form
\cite{Draine13}. The unaccounted oxygen is less readily explained \cite{Whittet10}. There is an
evident problem of oxygen budget in the dense interstellar medium, as the combined contribution of
gaseous CO and silicate/oxide dust are by far less of the lowest reference abundances
\cite{Asplund09}. Solid H$_2$O is an obvious O-bearing candidate. Astronomers have known for
decades that water is fairly common in the universe. Interstellar water ice, as distinct from
gaseous water molecules, was first identified toward the embedded protostellar Becklin-Neugebauer
object by Gillette \& Forrest (1973) \cite{Gillett73}. It is now widely detected in interstellar
dark clouds. These ices, composed mainly of water, also contain carbon monoxide and dioxide, simple
hydrides such as methane and ammonia, and a few other species \cite{boogert15}. Still, ice
contribution is not enough. According to Whittet (2010) \cite{Whittet10}, the most plausible
candidate for the depleted oxygen appears to be a form of O-bearing carbonaceous matter similar to
the organics found in cometary particles returned by the Stardust mission. Such materials may share
some similarities with refractory organic residues arising in the irradiation of icy mixtures by
ions \cite{baratta14} or X-rays \cite{ciaravella19}.

For its very nature, astronomical dust is likely to be an amalgam of a number of different
materials, very chaotic in composition and structure, with different individual substances
dominating at different wavelengths. These materials are thus fundamentally different from
terrestrial materials. Nuth \& Hecht (1990) \cite{Nuth90} introduced the concept of chaotic
silicates in which the level of disorder is even greater than for glasses, that are characterized
by the absence of long-range order in the atomic arrangement beyond nearest neighbours. Since
materials may be assembled in the agglomerate, an astronomical silicate cannot be considered a
solid with a definite stoichiometric composition. They may also occur in groups that recall solid
solutions, in which one or more types of atoms or molecules of the solid may be partly substituted
for the original atoms and molecules without changing the structure. Olivine and enstatite are
excellent examples of solid solutions. Forsterite, Mg$_2$SiO$_4$, and fayalite, Fe$_2$SiO$_4$, have
identical structures because the ions Mg$^{2+}$ and Fe$^{2+}$ are very nearly the same size and are
chemically similar. Very frequently amorphous silicates in space are misleadingly described in
terms of the optical properties of these materials. Indeed, as pointed out by Rietmeijer and Nuth
(2013) \cite{Rietmeijer13}, there are no amorphous silicates, as the word "silicate" already
implies that the material is crystalline and could be a mineral. Moreover, astronomical solids may
be porous and therefore of much lower density than a glass. Ultimately, the nature of an
astronomical silicate is rather loosely constrained, to same extent just limited to a material
whose infrared spectrum is dominated by Si-O stretching and bending vibrations. Thermal annealing
(e.g. in shocks) or intense X-ray irradiation \cite{Ciaravella16} of precursor materials, that were
probably amorphous, may explain the presence of crystalline silicates (see however Ritmejer and
Nuth 2013 \cite{Rietmeijer13}, for petrologic constraints) in circumstellar regions and
protoplanetary disks \cite{juhasz10}.

Even carbonaceous materials in space are difficult to constrain. A striking example is given by the
nature of the carrier of the interstellar ultraviolet extinction bump at 217.5 nm, that was
originally attributed to small crystalline graphite particles \cite{Stecher65}, followed by a
plethora of proposals including mixture of spheres composed of graphite, amorphous carbon, and
silicate \cite{Mathis89M}, irregular or fractal arrangement of graphite and amorphous carbon
\cite{Wright87}, polycyclic aromatic hydrocarbons \cite{Joblin92}, natural coal \cite{Papoular96},
and even electronic transitions of OH$^{-}$ ions in sites of low coordination in silicates
\cite{Steel87}. In general, carbonaceous materials contain greater or lesser hydrogen fractions,
varying proportions of different chemical bonding, and different degrees of long-range order. All
these forms of carbon can, under suitable conditions, be readily converted from one to another. The
manifold of possible bonding arrangements produces several allotropes of carbon of which the best
known are graphite, diamond and amorphous carbon. The physical properties of carbon vary widely
with the allotropic form. Amorphous carbonaceous materials cover a wide range of compositions, from
wide band gap, H-rich, aliphatic-rich a-C:H to narrow band gap, H-poor, aromatic-rich a-C
materials. The properties of a-C:H materials are determined by the $sp^3/sp^2$ ratio for the carbon
atoms and the hydrogen concentration. A C-H bond contributes to the formation of $sp^3$ bonding and
the reduction of the defects in the amorphous carbon network. In general, it is found that the
optical energy gap increases with hydrogen concentration \cite{Higa06}.

Remote observations such as extinction profiles were at one time the only sources of information
about interstellar dust. However, even if challenging it is now also possible to study in situ
interstellar grains entering the heliosphere by spacecrafts, and to collect them using
stratospheric aircrafts and balloons, and  satellite probes. Rare interstellar dust grains can now
be examined in the laboratory being returned by the Stardust/NASA mission. During the Cassini
mission around Saturn, millions of dust particles have been analyzed and now, thanks to the Cosmic
Dust Analyzer on board the probe, 36 grains of dust supposed coming from outside our solar system
have been detected \cite{altobelli16}. Results from different missions generally deviate one from
another, and typically they are not very consistent with the dust picture obtained by remote
observations \cite{Draine09}. Obviously, this is a field in which much work is required.

\begin{figure}
  \centering
  \resizebox{\textwidth}{!}{%
  \includegraphics{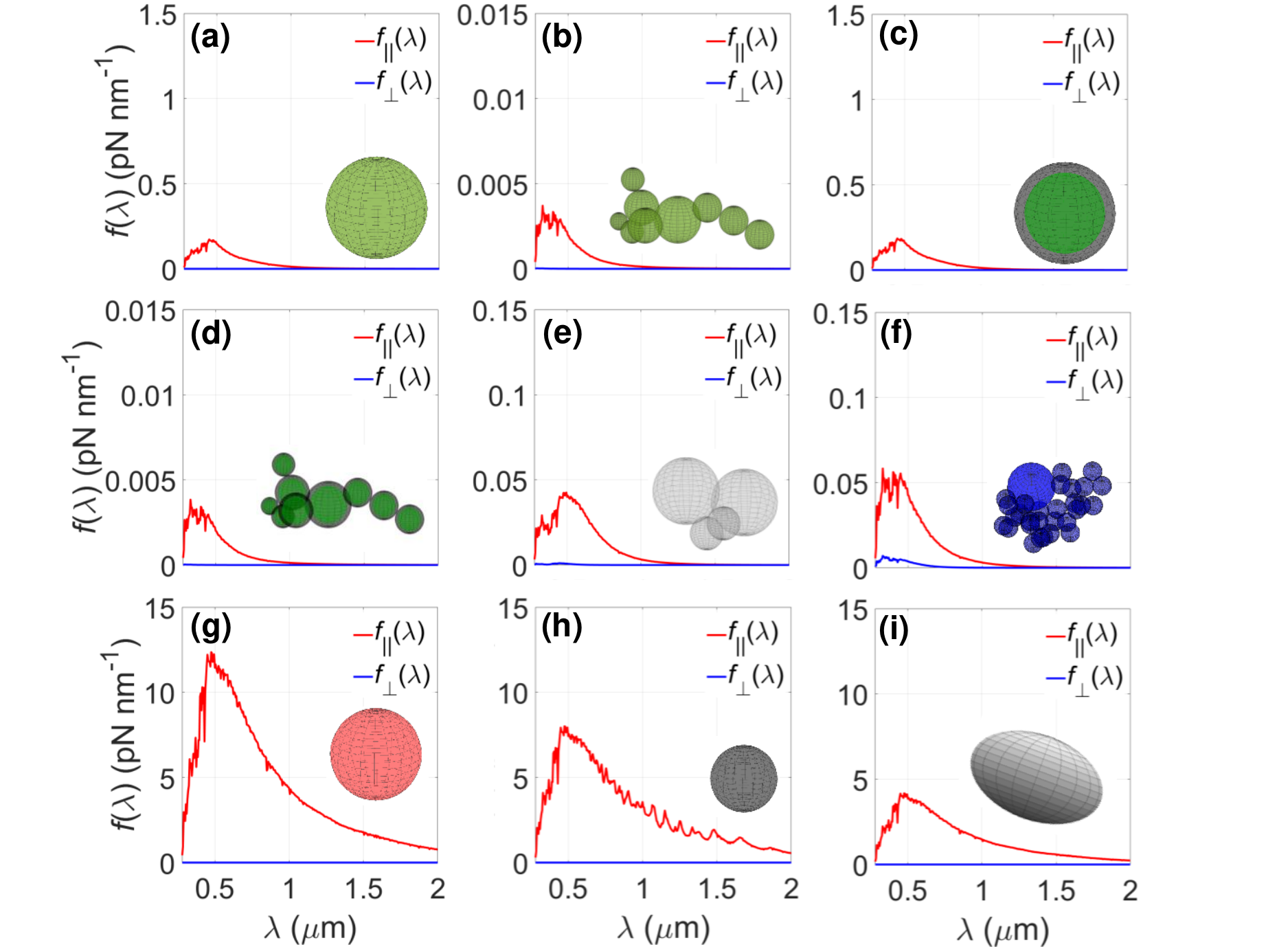}
  }
  \caption{Optical forces distributions exerted by the Sun on dust particles modeled as shown in Fig.~\ref{figure1}.
  The optical force parallel, $f_\parallel(\lambda)$ (blue line), and perpendicular, $f_\perp(\lambda)$ (red line),
  are intended respect to the light propagation direction $\hat{\mathbf{k}}$. In (a-d) we show calculations on olivine-aliphatic
  carbon structures. In (a) and (b), the constituents refractive indexes are mixed according to the Bruggeman
  criterion \cite{bohren2008absorption}. While in (c) and (d) the olivine and the carbon are distributed in a core-shell structure.
  In (e) we show results for the quenched melt silica particle, in (f) for the bunch-of-grape carbon Ca-rich, in (g) for the 'TP2' sphere, in (h) for the martian hematite sphere, and in (i) for the fassaite ellipsoid.}\label{figure2}
\end{figure}

\subsection{The interplanetary dust complex}
The interplanetary space of the solar system is very dusty, appreciable to the naked eye through
the faint solar colour cone of light above the western horizon after sunset, or above the eastern
horizon just before sunrise: the Zodiacal Light, a huge amount of fine dust particles that scatter
solar radiation in the interplanetary space. The brightness of the Zodiacal Light provides
information on the overall space density of the dust known as the interplanetary dust complex.

The interplanetary dust complex consists of microscopic (mainly rocky) particles, typically less
than a few millimetres in size, called micrometeoroids, moving in the interplanetary space of the
solar system \cite{rubin2010meteorite}. Dust is produced by collisions among solid bodies, by
disruptions of icy bodies \cite{nesvorny2006physical,flynn2009dust} and by cometary activity
\cite{fulle2020comets,guttler2019synthesis}. As such the interplanetary dust complex is an
inventory of the constituent materials of a large variety of solid bodies in the solar system.
Their cosmochemical study thus allows investigation of great diversity of astrogeobiological
processes occurred on their parent bodies. Remarkably, these include the most primitive bodies not
further processed by planetesimal accretion after aggregation in the protoplanetary disk
\cite{blum2017evidence}. In addition, since the bulk of extraterrestrial matter accreted by the
Earth is in the form of cosmic dust from the interplanetary dust complex, understanding its
composition provides also clues on outstanding issues like the contribution of extraterrestrial
matter to the Earth's geochemical budget  \cite{peucker1996accretion}, including its bearing on the
emergence of life \cite{sandford2016organic}. Based on particle-impact detection at the Long
Duration Exposure Facility/NASA, the main mass fraction of the submillimetre dust particles in the
Zodiacal cloud in the vicinity of the Earth have characteristic diameters of $\sim$100-200 $\mu$m,
and the flux of micrometeoroids entering the Earth's atmosphere is 40000$\pm$20000 tonnes per year
\cite{love1993direct}. Gr\"un et al. \cite{grun1985collisional} reached a similar conclusion and
quantified the spatial mass of cosmic debris at 1 AU as $10^{-16}$ g m$^{-3}$, with the largest
fraction in the $10^{-6}$ to $10^{-4}$ g mass range, based on the lunar microcrater record and
spacecraft micrometeoroid detectors.

How much asteroid collisions versus cometary activity contribute to the interplanetary dust complex
is an outstanding issue. Matter is complicated since such contributions are expected to vary
through the solar system's history; this is due to the stochastic fluctuations in the number of
mutual collisions in the asteroid belt and the number and level of activity of comets entering the
inner solar system. If astronomical observations  carried out by the Infrared Astronomical
Satellite (IRAS) and Cosmic Background Explorer (COBE) satellites in the 1980's indicated that
collisions in the Main Asteroid Belt (MAB) are the dominant source of dust in the near-Earth space
\cite{kortenkamp2001sources}, dynamical simulations predict in turn that the bulk of the Zodiacal
cloud can be best produced by debris derived from Jupiter Family Comets (JFCs) through spontaneous
disruption \cite{nesvorny2010cometary}. Oxygen isotopic data from relatively large micrometeorites
in the 100-1000 $\mu$m size range collected at the Earth's surface
\cite{cordier2014oxygen,suttle2020extraterrestrial,suttle2020isotopic} indicates that the
interplanetary dust complex is dominated by dust produced by cometary activity and by collisions
between primitive hydrous asteroids of carbonaceous chondrite compositions, with a subordinate
contribution from more evolved anhydrous asteroids (mainly ordinary chondritic and Vesta-like). An
additional small fraction of micrometeorites, with heavy oxygen isotope composition, may sample an
unknown body in the solar system, namely a body not yet sampled by macroscopic meteorites. The
isotopic statistics shows that primitive, volatile-rich material dominates the small size fraction,
whereas evolved anhydrous material dominates the large size fraction of the dust complex.

\paragraph{Physical and compositional properties of the interplanetary dust complex.}
Most of the compositional properties of the interplanetary dust complex derives from the
cosmochemical analyses of samples recovered from the Earth's: 1) surface, e.g., micrometeorites
collected in Antarctica \cite{cordier2014oxygen,dionnet2020x,suttle2020extraterrestrial}; 2)
stratosphere, by balloon born instruments as, e.g., DUSTER, designed for non-destructive and
uncontaminated collection of solid particles from tens of microns down to 200 nm in size
\cite{della2012situ} and by stratospheric NASA/aircraft passive sticking on silicon oil coated
plates \cite{rietmeijer2002earliest}. A critical contribution is also given by laboratory analyses
of samples, i.e., collected, and brought back to Earth, from asteroid surfaces
\cite{dionnet2020combining} and in a cometary coma \cite{rotundi2008combined,rotundi2014two}. In
addition, cometary dust was studied in situ from the onset of cometary activity to its cessation
after perihelion by the Rosetta/ESA space mission
\cite{della201667p,fulle2015density,rotundi2015dust}.

The pristine mineralogy of micrometeorites is best unravelled by the study of a specific class: the
unmelted, i.e., particles that did not experienced alteration by frictional heating during
atmospheric passage \cite{luigi15micrometeorites}. Based on petrographic and geochemical data, most
of the micrometeorite flux reaching Earth today is dominated (>50\% of unmelted micrometeorites
across all size fractions) by fine-grained and hydrated carbonaceous chondrite material affine to
the CM, CR matrices and CI chondrites, with some specimens similar to the ungrouped meteorite
Tagish Lake. They are therefore a major component of the near-Earth dust complex in agreement with
isotopic statistics \cite{cordier2014oxygen}. Coarse-grained micrometeorites, which are primarily
fragments of chondrule, and CAI (Ca-, Al-rich inclusions) represent a smaller component (<25\%) of
the micrometeorite flux \cite{van2012chondritic,suttle2020extraterrestrial}. Thus, the major
constituent minerals are olivine, low-Ca pyroxene, magnetite, sulphides, metal and hydrous Fe-Mg
silicates like serpentine and saponite. The density of unmelted micrometeorites varies greatly
according to their mineral composition. Mean density for unmelted fine-grained micrometeorites is
$\sim$1.4 g cm$^{-3}$ (average data from Kohout et al. \cite{kohout2014density}), but can be more
than twice for coarse-grained micrometeorites. Micrometeorites range in size from 10 $\mu$m to 2000
$\mu$m \cite{rubin2010meteorite}, however particles up to 3000 $\mu$m in size have been reported in
the Transantarctic Mountains micrometeorite collection
\cite{rochette2008micrometeorites,suavet2009statistical}, and their mass varies within the nanogram
to the milligram range. A statistically significant number of micrometeorites in the 200-700 $\mu$m
size range shows that the micrometeorite size distribution is bimodal, with peaks centred at
$\sim$145 and $\sim$250 $\mu$m. This suggests that the micrometeorite flux is composed of multiple
dust sources with distinct size distributions: fine-grained material from primitive objects and
coarse-grained material from evolved bodies \cite{suttle2020extraterrestrial}.

DUSTER particles, collected with a strict contamination protocol during different stratospheric
balloon flight campaigns at 30-40 km altitudes, are the smallest (down to 200 nm) meteoritic
objects available for laboratory investigation (see Fig. \ref{figure1}f,h,l). These are the
residues of porous, structurally weak bolides crossing the Earth's atmosphere. They contribute to
the daily input of stratospheric extra-terrestrial material with: 1)tens of microns down to
sub-micrometre aggregates of nanometer-sized spherical grains, produced by vaporization and
quenching; 2) partially annealed minerals; and 3) unprocessed minerals
\cite{corte2013meteoric,rietmeijer2016laboratory}. DUSTER particles composition includes alumina,
aluminosilica, plagioclase, fassaite, silica, CaCO$_3$, CaO, extreme F-rich C-O-Ca particles, and
oxocarbon. These are particle linked to the friable CI and CM carbonaceous chondrite, and
unequilibrated ordinary chondrite meteors that are the most common source of bolides and fireballs
\cite{rietmeijer2016laboratory}.

Similar particles are collected at lower quotes in the stratosphere by NASA/aircrafts:
Interplanetary Dust Particles (IDPs), numerous and fine-grained, are available for laboratory
analyses \cite{rietmeijer2002earliest,rotundi2007combined}. They have a chondritic overall
composition, masses in the order of pico- to nano-grams and sizes of few microns
\cite{rubin2010meteorite}, but they can reach hundreds of microns for the Giant Cluster IDPs class.
They mainly consist of extremely fragile aggregate of crystalline and amorphous materials with
grain-size typically in the order of 100s of nm or smaller. Hydrous IDPs are mainly massive objects
with fibrous or platy surface textures. They consist mainly of hydrous mineral assemblages, mostly
smectite and lesser serpentine and silicate glass, with minor amounts of diopside, forsterite,
chromite, Fe- and Ni-sulfides, schreibersite (Ni-phosphide), magnetite, glasses of silicatic
composition and disordered carbonaceous material. Among IDPs are the aggregate anhydrous IDPs,
which are extremely fine-grained, highly porous (up to 70\%) and low-density (0.3-0.6 g cm$^{-3}$).
Following Rietmeijer's  classification\cite{rietmeijer2002earliest}, they consist of a matrix
aggregates of generally spherical entities, $\sim$0.1 to $\sim$1 micron in size, with embedded
variable amounts of $\sim$5 micron-sized Mg, Fe- and Ca, Mg, Fe-silicates, Ni-free and low-Ni
pyrrhotite, iron oxides, and amorphous materials. Aggregate IDPs are supposed to have a cometary
origin \cite{brunetto2011mid} testified also by the comparison with cometary dust particles
returned by the Stardust/NASA space probe \cite{rotundi2008combined,rotundi2014two}. In fact, dust
particles after ejection from the nucleus form a dust flux \cite{della2019giada}, which undergoes
to solar radiation pressure effects, as observed in situ by GIADA onboard the ESA Rosetta space
probe \cite{della2016giada}, to gas drag and to gravity force \cite{zakharov2018asymptotics}. These
dust particles either remain tied to the nucleus or they are liberated into space feeding the
interplanetary dust complex, where their orbital evolution is mainly controlled by radiation
pressure and the Poynting-Robertson light drag \cite{dermott2001orbital}. This causes particles
from 1 to 1000s of $\mu$m in size to spiral slowly into the Sun, i.e., to gradually reduce the
orbit's eccentricity and semi-major axis, in a $10^4$ to $10^6$ year time scales, intercepting in
some cases Earth's atmosphere, where they are collected as IDPs.

\subsection{Planetary dust}
Many terrestrial planets and satellites of both terrestrial and giant planets in our solar system
show dusty environmental conditions. In particular, the mostly explored dusty bodies are the Moon
and Mars.

\paragraph{The Moon.}
The Moon is the only Earth's natural satellite. Since the Moon has neither a magnetic field nor a
significant atmosphere, the lunar regolith and the near-surface environment are mainly affected by
space weathering processes such as meteoroid impacts, solar ultraviolet radiation, solar wind,
galactic cosmic rays and plasma processes in the tail of the Earth's magnetosphere.

On the Moon, all the locations explored so far have cratered surfaces covered with loose regolith
of several meters. We can therefore assume that the entire surface is covered by regolith although
the thickness may vary (about 5 m in the maria, up to 20 m in the highlands). Such regolith layer
is a cohesive, dark grey to light grey, very-fine-grained, clastic material consisting of a mixture
of a wide variety of materials including fragments of highland anorthosites, some forms of KREEP
volcanic rocks, mare basalts and volcanic glass, plus a small meteoritic component
\cite{lucey2006understanding}. The influence of the solar wind and high-energy particles of solar
and cosmic origins induces the implantation of H, He, and many rare elements into the regolith.
Continued reworking by micrometeoroids of the hydrogen-enriched regolith particles causes the
material melting and the reactions (particularly, H with FeO), producing water vapor and submicron
grains of metallic iron sintered into glass \cite{mckay1991lunar}.

Considering the samples returned by the Apollo and Luna missions, the fine fraction of the lunar
regolith (45-100 $\mu$m in mean size) comprises about a half of the lunar regolith by weight; the
superfine fraction (particles smaller than 10 $\mu$m) about 10\%, while the particles smaller than
2 $\mu$m, the finest regolith fraction, make up 1-2\% of the mass. The shape of dust particles is
extremely irregular and highly variable, ranging from angular with sharp edges to spheroidal. The
density of individual particles is usually assumed in the range from 2.7 to 3.0 g cm$^{-3}$
\cite{carrier1991physical}. Individual particles may be glass-bounded aggregates
\cite{mckay1991lunar} called agglutinates. Agglutinate particles may have lower density (higher
porosity), they are usually <1 mm and contain minute droplets of Fe metal (much of which is very
fine-grained, single domain FeO), and troilite (FeS). They formed by the melting and mixing
produced by micrometeoritic bombardment of the lunar regolith.

The optical parameters of the dust particles (\textit{i.e.}, the real and imaginary parts of the
complex refractive index) depend on the composition and can vary significantly, not only over a
wide region, but even locally in a microscopic scale. For the glassy component, the real part
values range from 1.570 to 1.749 and they vary directly with the total Fe and Ti contents and
inversely with the Al content \cite{chao1970lunar}. The imaginary part of the complex refractive
index is more variable and can span from 0.0005 up to 0.15 according to the composition and
wavelength of interest.

Electrical properties of the dust particles are characterized by the extremely low electrical
conductivity (approximately $10^{-14}$ ohm/m in shadowed areas \cite{carrier1991physical}, but
about a hundred times higher under solar-light exposure) permitting accumulation of electrostatic
charge under ultraviolet irradiation.

\paragraph{Mars.}
Mars, the fourth planet of our solar system, has many features in common with the Earth. Much of
the Martian surface is covered by unconsolidated soils (dust) derived from impact, aeolian and
other sedimentary processes. Such dust is likely produced by impacts early in Mars' history and
subsequently recycled at its surface. Sometimes the wind in Mars' thin atmosphere blows the dust on
the surface into dust storms, carrying dust particles up to altitudes of about 50 km. The
composition of the Martian dust grains was obtained from several space missions, starting from the
mission Mariner 9 in 1971 \cite{toon1977physical}.

It is possible to build a mineralogical model of the dust using all the data acquired from the
Mariner 9 and Viking Landers \cite{clark1982chemical} space missions. Best fit results are those
obtained by mixing different kinds of clay, with montmorillonite Si$_2$Al$_4$O$_{10}$(OH)$_2$
$\cdot$nH$_2$O as a base, mixed with nontronite (where part of Al is replaced by Fe$^{3+}$) and
saponite (where Al is replaced by Mg). Indeed, data by the X spectrometer aboard the Pathfinder
showed higher abundances of magnesium and aluminum, and lower abundances of iron, chlorine and
sulphur \cite{foley2003final} respect to what expected. The rocks analyzed were similar to
andesites, terrestrial materials derived from the fractionation of basalt in intrusive conditions.

Data from recent rovers (e.g. MSL Curiosity at Rocknest, Oct 2012) and orbital spacecraft show that
Martian surface is dominated by a soil (dust grain dimension <150 $\mu$m) of basaltic composition
with  primarily pyroxene, plagioclase feldspar, and olivine, as well as minor amounts of Fe and Ti
oxides (e.g., magnetite, ilmenite, and hematite) and alteration minerals (e.g., sulfates,
phyllosilicates, and carbonates)
\cite{morris2006,yen2005integrated,mcsween2010determining,bish2013x}. Dust particles properties
changes during dust storms \cite{chen2020dust}. In general, dust particle sizes range from 1.2 to
about 4.1 $\mu$m with a mean dust radius of about 1.6-1.8 $\mu$m
\cite{oliva2018properties,wolff2009wavelength} and a positive correlation between dust opacity and
particle sizes.

Phase function results show asymmetry parameter values of $g = 0.601$ $\pm$ 0.108 for high
atmospheric dust loading scenarios and $g = 0.710$ $\pm$ 0.065 for non-dusty periods. Regarding the
shape of the particles, considering a modified log-normal aspect ratio distribution for a mixture
of spheroids, data suggest more elongated particles are present during dust storms, with aspect
ratios of 2.8 $\pm$ 0.9 for high-opacity days, in contrast to values of 1.8 measured during
post-storm period. The particles single scattering albedo is found ranging in the solar band
between 0.89-0.90 (dark regions) and 0.92-0.94 (bright regions) \cite{wolff2010ultraviolet}.

Moreover, there are a very strong and a fast feedback between sand/dust emission and electric field
enhancement and a linear relation between the concentration of lifted dust and the generated
electric field \cite{esposito2016role}.

\begin{figure}
  \centering
  \resizebox{0.93\textwidth}{!}{%
  \includegraphics{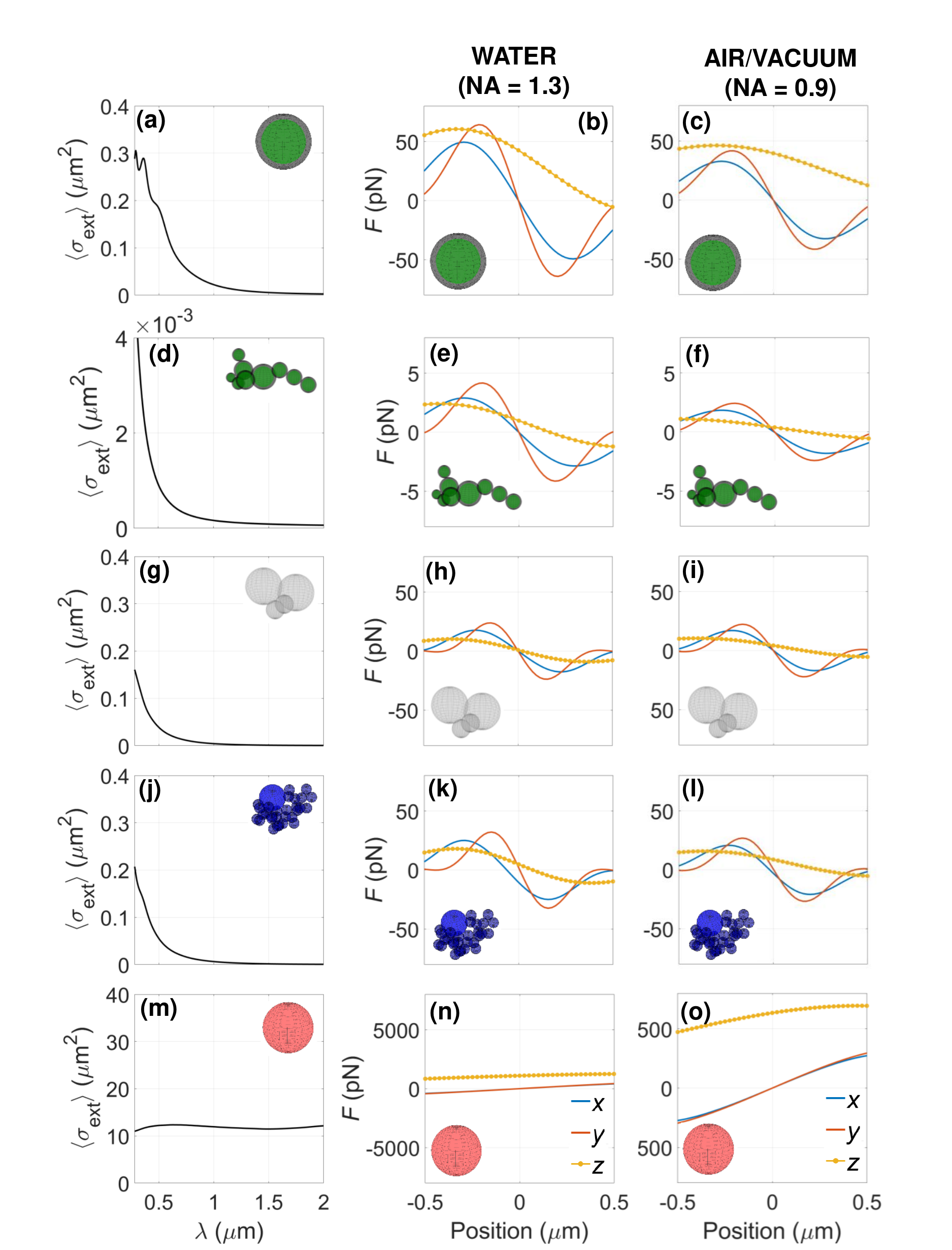}
  }
  \caption{Extinction cross-sections, $\langle \sigma_\mathrm{ext} \rangle$, and optical trapping force, $F$, components along $x$ (blue line), $y$ (red line), and $z$ (yellow dots) when the particle is trapped in water (center column) or in air (right column). Results in (a-c) are related to the olivine-aliphatic carbon core-shell sphere. (d-f) show calculations for the olivine-aliphatic carbon core-shell cluster. (g-i) concern the quenched melt silica. (j-l) show the results for the bunch-of-grape carbon Ca-rich. (m-o) are the results for the 'TP2' sphere. For the optical trapping calculations the laser power is fixed at 50 mW and the wavelength at $0.83$ $\mu$m.}\label{figure3}
\end{figure}

\section{Methods and models for space tweezers calculations}\label{sect:Model and methods}
Optical forces and optical trapping are the consequence of the electromagnetic momentum
conservation during a light scattering process \cite{jones2015optical,borghese2007scattering}. To
understand theoretically how  light interacts mechanically with matter, Maxwell's equations and the
integration of the averaged Maxwell stress tensor have to be performed
\cite{mishchenko2002scattering,borghese2007scattering}. However, such calculations can be
computationally complex, $e.g.$ when we deal with non-spherical or non-homogeneous particles, and
the use of approximations can be also advantageous \cite{jones2015optical,polimeno2019optical}. The
range of validity of the different approximations is determined by the size parameter $x =
k_\mathrm{m} r$, where $k_\mathrm{m} = 2 \pi n_\mathrm{m} /\lambda_0$ is the radiation wavenumber
in the medium surrounding the particle, $r$ is the particle size, $\lambda_0$ is the wavelength in
vacuum, and $n_\mathrm{m}$ is the medium refractive index \cite{jones2015optical}. When $x \ll 1$,
the particle size is much smaller than the wavelength, and the dipole approximation can be adopted,
treating the particle as a dipole \cite{chaumet2000time,arias2003optical,albaladejo2009scattering}.
In the opposite case, when $x \gg 1$, the ray optics regime appears computationally the most
suitable, and the optical fields are simply represented as a collection of $N$ light rays according
to the geometrical optics \cite{ashkin1992forces,jones2015optical}. Each ray has a $N$-th portion
of the total incident power $P_\mathrm{i}$, and a linear momentum per second $n_\mathrm{m}
P_{\mathrm{i},N}/c$ \cite{born1999principles}. According to the Snell's law, a single ray which
impinges on a surface with an certain incident angle is partly reflected and partly transmitted.
Therefore, the ray changes its direction and, consequently, its momentum causing a reaction force
on the center of mass of the particle \cite{callegari2015computational}.

In the intermediate regime, that is when the particle size is comparable with the light wavelength
($x \simeq 1$) or for non-spherical and non-homogeneous particles like cosmic dust, we need to
resort to full electromagnetic theory and the T-Matrix formalism has proved to be computationally
advantageous \cite{waterman1971symmetry,borghese1980addition,borghese2007scattering}. Any particle
can be modeled through a sphere, clusters or aggregates of spheres, spheres with spherical
(eccentric) inclusions, and multilayered spheres \cite{borghese2007scattering}. This formalism
consists in the expansion of the electromagnetic fields into a basis of vector spherical harmonics
applying the boundary conditions across the particle surface
\cite{mishchenko2002scattering,borghese2007scattering}. The incident, and the internal fields
inside the particle have to be regular at the origin while the scattered field is such as to
satisfy the radiation condition at infinity
\cite{borghese2007scattering,fucile1997optical,ishimaru1991wave}. The transition matrix,
$\mathbb{T}$, connects the amplitudes of the scattered fields to the amplitudes of the incident
ones upon a multipole expansion of the fields. Its order is, in principle, infinite and thus it
must be truncated to a multipole index value $L_M$ \cite{borghese2007scattering}. Such value is
chosen to ensure the required accuracy of the transition matrix elements. However, for a cluster of
$N$ spheres this implies the solution of a system of order $D_M = 2 N L_M (L_M + 2)$, which may
become too large \cite{borghese2007scattering}. Actually, the inversion of the scattering matrix is
responsible for most of the computation time required for the calculation that scales as $D_M^3$.
Thus, on account of the definition of $D_M$, the computation time scales as $L_M^6$ whereas the
storage requirements scale as $L_M^4$, so that it pays, in terms of both CPU time and storage
requirements, to keep $L_M$ as low as practicable. Thus, the choice of an appropriate value of
$L_M$, in order to satisfy computation time and storage requirements, is of crucial importance
\cite{iati2004optical,saija2003efficient}.

The diversity of dust particles in an astrophysical context implies a richness of models that we
need to build in order to calculate realistic optical forces for space tweezers applications. Here,
we consider several models of extraterrestrial dusts with shape and composition inspired by
interstellar particles, DUSTER samples \cite{corte2013meteoric,rietmeijer2016laboratory}, and Moon
or Mars dust analogs. We show these models in Fig.~\ref{figure1}. On the top row, the homogeneous
and stratified single/aggregated spheres emulate hypothetical interstellar dust grains whose
constituents are olivine and aliphatic carbon \cite{antonia2008stratified}. The olivine refractive
index is provided by Draine \& Li while the carbon one by Ashok et al.
\cite{draine1984optical,palik1998handbook}. In Figs.~\ref{figure1}a and \ref{figure1}b, the
constituents refractive indexes are mixed in such a way as to treat the particle homogeneously with
a single effective refractive index according to the Bruggeman criterion
\cite{bohren2008absorption}. On the other hand, in Figs.~\ref{figure1}c and \ref{figure1}d, an
olivine core is considered covered by a carbon layer \cite{antonia2008stratified}. The spheres
radius of Figs.~\ref{figure1}a and \ref{figure1}c is $r=0.15$ $\mu$m. On the other hand, the
clusters of Figs.~\ref{figure1}b and \ref{figure1}d are composed by $9$ spheres of different sizes
with the major semi-axis $r=0.16$ $\mu$m. In Fig.~\ref{figure1}e, we present a model according to a
Field Emission Scanning Electron Microscope (FESEM) image of a silica particle clustering arranged
in quenched melt spheres, shown in Fig.~\ref{figure1}f, and collected by DUSTER
\cite{rietmeijer2016laboratory}. The refractive index is provided by Malitson
\cite{malitson1965interspecimen}. The model is composed by 4 spheres of different radius with the
major semi-axis $r=0.23$ $\mu$m. Fig.~\ref{figure1}g represents the model of condensed Ca[O]
nanograins that are accreted onto a larger melted aggregate of tiny carbonate grains, shown in the
FESEM image of Fig.~\ref{figure1}h, and collected by DUSTER
\cite{corte2013meteoric,rietmeijer2016laboratory}. The larger sphere is calcite and the other
spheres are CaO. The cluster model is composed by 30 spheres with the calcite refractive index
provided by Ghosh while the calcium oxide one is provided by Liu \& Sieckmann
\cite{ghosh1999dispersion,liu1966refractive}. Moreover, its major semi-axis has $r=0.25$ $\mu$m. In
Fig.~\ref{figure1}i, a spherical model ($r=1.25$ $\mu$m) of the particle Fe, Mg-rich TP2, collected
by DUSTER, is shown \cite{rotundi2014two}. We consider an effective refractive index obtained
mixing iron (67 \%) and magnesium (33 \%), according to the Bruggeman criterion whose the two
refractive indexes are respectively provided by Johnson \& Christy and Hagemann
\cite{johnson1972optical,hagemann1974desy,bohren2008absorption}. In Fig.~\ref{figure1}j, a
spherical model of Martian hematite ($r = 1$ $\mu$m) \cite{oliva2019database}. In
Fig.~\ref{figure1}k, a spherical model of Lunar regolith ($r = 1$ $\mu$m) \cite{egan1973optical}.
Fig.~\ref{figure1}l shows the model of a microscale fassaite ellipsoidal, collected by DUSTER, and
shown in a FESEM image of Fig.~\ref{figure1}m \cite{rietmeijer2016laboratory}. We consider an
effective refractive index constructed mixing silica (65 \%), and CaO (35 \%) according to the
Bruggeman criterion \cite{bohren2008absorption}, the major semi-axis is 2 $\mu$m. All non-spherical
models are oriented in such a way that their major axis is aligned with the incident light
propagation direction.
\begin{table}
  \centering
  \resizebox{0.9\textwidth}{!}{%
  \includegraphics{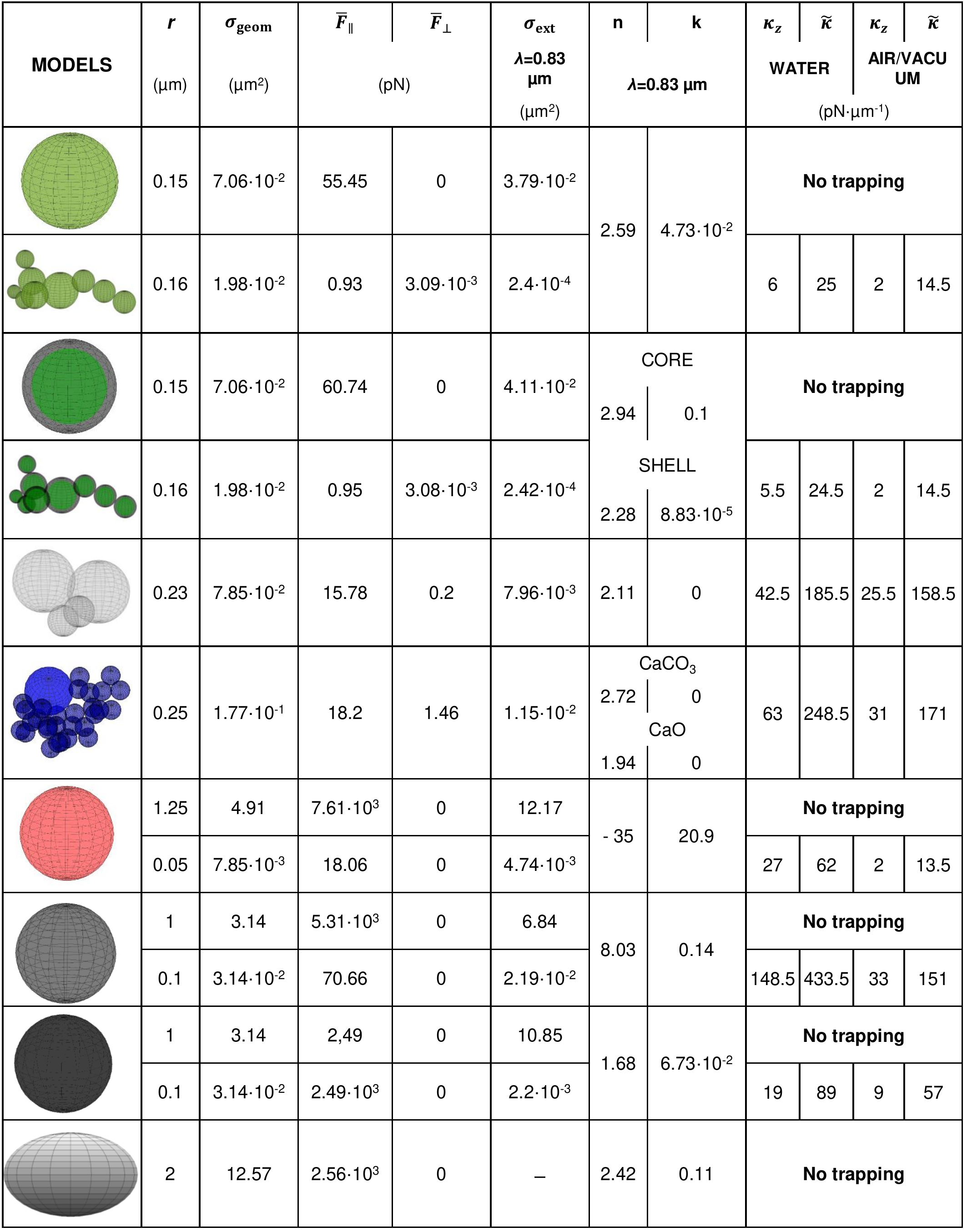}
  }
  \caption{Summary table of radiation pressure and optical trapping properties. For each model particle we show: the radius, $r$, or the radius of the smallest sphere enclosing the cluster; the geometric cross-section, $\sigma_\mathrm{geom}$; the integrated optical force on the solar intensity spectrum along the parallel, $\bar{F}_\parallel$, and perpendicular direction, $\bar{F}_\perp$, with respect to the light propagation; the extinction cross-section $\sigma_\mathrm{ext}$ at the trapping wavelength $\lambda=0.83$ $\mu$m; the real, $n$, and the imaginary part, $k$, of the refractive index at the trapping wavelength $\lambda=0.83$ $\mu$m; and the optical trap stiffnesses along the optical axis, $\kappa_z$, and perpendicular to it, $\tilde{\kappa}$, in water and in air.}\label{figure4}
\end{table}

The calculation of the solar radiation pressure and optical trapping forces on model particles of
Figs.~\ref{figure1}a - \ref{figure1}k is carried out with the T-matrix formalism because their size
parameters fall within the range $x \approx [0.1 - 6]$. On the other hand, when the particle size
parameter is too high (e.g. $x\geq 10$) such as in the fassaite micro ellipsoid of
Fig.~\ref{figure1}l, the calculation is carried out in the ray optics approximation. For this case
we exploited the optimized computational MATLAB codes for dielectric particles provided by
Callegari et al. \cite{callegari2015computational}
(http://opticaltweezers.org/software/otgo-optical-tweezers-geometrical-optics/).

\section{Results on solar radiation pressure and optical trapping of dust particles}\label{sect:Results and discussion}
Solar radiation pressure calculations are important to understand its relevance in optical trapping
applications in space. In general external forces, such as solar radiation pressure, can have
detrimental effects on optical trapping of particles in space or the high atmosphere. Here we aim
to show that T-matrix methods can be used to give accurate estimates of these effects on individual
dust particles. The total radiation force, $\mathbf{\bar{F}}$, and force spectrum, $f_s (\lambda)$,
that the Sun exerts on particles are calculated as:
\begin{equation}\label{eq:solar_force_integration}
   \bar{F}_s = \int_{\lambda} \mathrm{d}\lambda \ f_s (\lambda) \ ,
\end{equation}
\begin{equation}\label{eq:solar_force_distribution}
f_s (\lambda) = \frac{i_\odot(\lambda)}{I_\odot} \ F_{\mathrm{rad},s} (\lambda) \ ,
\end{equation}

\noindent where the index $s=(\parallel,\perp)$ specifies respectively the parallel and the
orthogonal component of $F_{\mathrm{rad},s} (\lambda)$ respect to the radiation incident direction,
$\hat{\mathbf{k}}$. The expressions of $F_{\mathrm{rad},s}(\lambda)$ are the wavelength dependent
optical forces calculated in the ray optics approximation \cite{jones2015optical} or in the
T-matrix approach \cite{borghese2007scattering} according to the different models under
investigation. The term $f_s (\lambda)$ specifies the spectral force distribution obtained scaling
the computation outputs $F_{\mathrm{rad},s} (\lambda)$ upon the solar radiation intensity
distribution $i_\odot(\lambda)$ normalized at the solar irradiance $I_\odot$, $i.e.$, $I_\odot =
\int_{\lambda} \mathrm{d}\lambda \ i_\odot(\lambda) = 1.34$ kW$/$m$^2$ \cite{standard2012g173}.
Therefore, the considered spectrum is chosen in the range $\lambda = [0.28 - 2]$ $\mu$m that is
when the light emission by the Sun has the maximum intensity.

In the Fig.~\ref{figure2}, we plot the solar radiation force distributions $f_s (\lambda)$
(Eq.~\ref{eq:solar_force_distribution}) for all the models presented in  Fig.~\ref{figure1}. In
particular, we show the force, $f_\parallel(\lambda)$, parallel to the propagation direction as a
red line, and the perpendicular one, $f_\perp(\lambda)$, as a blue line. Figures.~\ref{figure2}a-d
concern the calculation on different models for the olivine-aliphatic carbon structures.
Figure~\ref{figure2}e is referred to the quenched melt silica, Fig.~\ref{figure2}f the
bunch-of-grape carbon Ca-rich, Fig.~\ref{figure2}g the 'TP2' sphere, and Fig.~\ref{figure2}h the
martian hematite sphere. Finally, Fig.~\ref{figure2}i shows the fassaite ellipsoid for which the
radiation force is calculated with ray optics. We note how the parallel component, $f_\parallel$,
is much larger than the perpendicular one, $f_\perp$. In fact, the parallel component is
proportional to the particle extinction cross section, while the transverse one is related to the
asymmetry parameters, $g_i$, related to the non-sphericity of the scatterer
\cite{mishchenko2001radiation,borghese2007scattering,saija2005transverse}. Thus, for cylindrically
symmetric particles the component $f_\perp$ is zero (Figs.~\ref{figure2}a, \ref{figure2}c,
\ref{figure2}g, \ref{figure2}h, \ref{figure2}i). Figures.~\ref{figure2}b, \ref{figure2}d, and
\ref{figure2}e show results for elongated nanoscale clusters that generally align with the incident
electric field direction \cite{borghese2008radiation} yielding a low value of $f_\perp$. In
Fig.~\ref{figure2}f we report the highly non-symmetrical model of the bunch-of-grape carbon Ca-rich
for which $f_\perp$ is quite strong and $\bar{F}_\perp \simeq 1.46$ pN, a value comparable to
$\bar{F}_\parallel \simeq 18.2$ pN (Tab.~\ref{figure4}). In summary, the parallel component of the
solar radiation pressure, $f_\parallel$,  describes an optomechanical interaction of the solar
radiation pressure with the extraterrestrial dust models in the tens of piconewton range. On the
other hand, for non-spherical particles the perpendicular component of the solar radiation pressure
can drive more complex transverse or rotational dynamics \cite{klavcka2001motion}.

We now focus on the systematic characterization of optical trapping forces in optical tweezers,
$i.e$, a single Gaussian beam focused by a high-NA objective. In our calculations we fix the laser
wavelength at $0.83$ $\mu$m, that is a typical wavelength for optical tweezers experiments in the
near-infrared, and the power $P = 50$ mW. In Fig.~\ref{figure3}, we show the three Cartesian
components of the trapping force in the neighborhood of the optical tweezers paraxial nominal focus
placed at the origin of the coordinate system ($x=y=z=0$). The trapping position of the particle in
the axial $z$-direction does not typically coincide with the origin because of the 'pushing' effect
of the optical scattering force. To calculate the force on the particle at the equilibrium position
$C_\mathrm{eq} = (x_\mathrm{eq}, y_\mathrm{eq}, z_\mathrm{eq})$, the $z$ axial coordinate at which
the axial force vanishes must firstly be found. Hence, the force plots in the transverse directions
$(x, y)$ can then be calculated. In the left column of the Fig.~\ref{figure3}, we present the
extinction cross-sections $\sigma_\mathrm{ext}$ for the visible and near-infrared wavelength
spectrum. The extiction cross section, $\sigma_\mathrm{ext}$, takes into account the rate at which
the energy is removed from the electromagnetic wave through scattering and the absorption, allowing
us to understand how effectively trapping takes place \cite{jones2015optical}. The trapping arises
when the focused incoming field generates a restoring force proportional to the particle's
displacement from an equilibrium point, and that, for small displacements, behaves harmonically
\cite{Polimeno2018}. Therefore, trap stiffnesses are defined as:

\begin{equation}
   \kappa_i \equiv \frac{\mathrm{d}F_i}{\mathrm{d} x_i}\bigg|_{x_{\mathrm{eq},i}} \ .
\end{equation}

\noindent We calculate optical trapping in water ($n_{\rm m}=1.33$) with an objective NA $= 1.3$
(middle column of Fig.~\ref{figure3}) and in air or vacuum ($n_{\rm m}=1$) with NA $= 0.9$ (right
column of Fig.~\ref{figure3}). We note that here we focus only on electromagnetic calculations,
neglecting  thermal fluctuations and hydrodynamics effects due to the surrounding medium
\cite{volpe2013simulation}. Generally, for a given particle, optical trapping in water is
stabilized by the overdamped viscous dynamics in the fluid, while in air or vacuum the underdamped
dynamics might be more critical for stable optical trapping \cite{svak2018transverse}. Moreover,
the higher NA in water causes the equilibrium point $C_\mathrm{eq}$ to be closer to the nominal
focus than in air. This is confirmed by comparing the graphs in  Fig.~\ref{figure3}, central column
(in water), with those of the right column (in air). The dielectric particles, such as the quenched
melt silica (Figs.~\ref{figure3}h, \ref{figure3}i), and carbonate cluster (Figs.~\ref{figure3}k,
\ref{figure3}l) exhibit fairly stable trapping. Even the interstellar dust model of a sphere
aggregate  can be trapped thanks to the relatively low extinction (Figs.~\ref{figure3}e,
\ref{figure3}f). Indeed, for these model particles we are able to extrapolate the trapping
constants along the axial direction, $\kappa_z$, and along the perpendicular direction,
$\tilde{\kappa} \equiv (\kappa_x + \kappa_y)/2$ (Tab.~\ref{figure4}). On the other hand, model
particles, like the interstellar dust sphere (Figs.~\ref{figure3}b, \ref{figure3}c), or the Fe-Mg
sphere (Figs.~\ref{figure3}n, \ref{figure3}o), can not be trapped neither in water nor in air. A
similar behavior is also exhibited by the hematite sphere, regolith sphere, and the ellipsoidal
fassaite as shown in Tab.~\ref{figure4} by observing the corresponding trapping constants. The
reason is to be found in their large size (Fe-Mg sphere, hematite, regolith, and fassaite) and on
their strong absorption (interstellar dust, Fe-Mg sphere, hematite, regolith and fassaite). To
further confirm, we have calculated the trapping properties of the Fe-Mg, hematite spheres by
reducing their size such that $r_\mathrm{Fe-Mg} = 0.05$ $\mu$m and $r_\mathrm{hem} = 0.1$ $\mu$m.
In this way, the trap stiffnesses can be extrapolated (Tab.~\ref{figure4}) and a quantitative
indication of $\sigma_\mathrm{ext}$ is provided in Tab.~\ref{figure4}. This behaviour is similar to
the optical trapping of metal nanoparticles that can be efficiently trapped at small size where
absorption and extinction cross sections are still small in the near-infrared
\cite{amendola2017surface}.

\section{Conclusions}
After a review on the role of interstellar, interplanetary, and planetary dust in the universe, we
studied computationally the solar radiation force and optical trapping properties for different
cosmic dust particles. We showed that the radiation force exerted by the Sun is not trivially
negligible: it can actively influence the dynamics of the modeled dust particles at the nano- and
microscale when compared to, \textit{e.g.}, optical trapping forces. Furthermore, we calculated
single-beam optical trapping properties for cosmic dust modeled particles both in water and in air
with parameters close to typical optical tweezers experiments in the near-infrared. We found that
dielectric and weakly absorbing particles can be captured, while microparticles with stronger
absorption show a scattering component of the optical force that prevents optical trapping in a
standard single-beam optical tweezers. This detrimental effects can be overcome through dual-beam
traps made by stationary counter-propagating laser beams
\cite{zemanek2002simplified,donato2018optical}.

This study opens perspectives for the application of optical tweezers techniques to solar system
study, $e.g.$, cometary particles analyses including the volatile component, dust particles in the
Martian atmosphere and/or on the Martian and Lunar surfaces. Such applications will also be
strategic within curator facilities designed for the uncontaminated handling and preliminary
characterization of extraterrestrial samples returned by space probes. The opportunity to apply
optical tweezers to planetary particulate matter in the frame of curation facilities can pave the
way for space applications for \textit{in situ} physical chemical analyses. The accurate knowledge
of the (optical, microchemical and electrical) properties of extraterrestrial dust allows the
design of an appropriate optical tweezers system for space applications, \textit{e.g.}, to be
mounted on stratospheric platforms, onboard future cometary probes, or landers/rovers to Mars or to
the Moon.

\section*{Acknowledgements}
We acknowledge financial contribution from the agreement ASI-INAF n.2018-16-HH.0, project ``SPACE
Tweezers''. M.A.I., D.B.C., M.G.D., A.F., P.G.G., and O.M.M. acknowledge support from the MSCA ITN
(ETN) project ``Active Matter'', project number 812780.

\bibliographystyle{unsrt}
\bibliography{biblio_radiation_force}
%
%
%
\end{document}